\documentclass[lettersize,journal]{IEEEtran}
\usepackage{amsmath,amsfonts}
\usepackage{enumitem}
\usepackage{algorithm}
\usepackage{array}
\usepackage[caption=false,font=normalsize,labelfont=sf,textfont=sf]{subfig}
\usepackage{textcomp}
\usepackage{stfloats}
\usepackage{url}
\usepackage{verbatim}
\usepackage{graphicx}
\usepackage{siunitx}
\usepackage{cite}
\usepackage{changepage} 
\usepackage{tikz}
\usepackage{pgfplots}
\usepackage{algpseudocode}
\usepackage{amssymb, mathtools}
\usepackage{lipsum}
\pgfplotsset{compat=1.17}
\usetikzlibrary{positioning,arrows.meta,calc,decorations.pathmorphing,shadows.blur,fit,patterns,backgrounds}
\begin{document}

\title{Two-Stage Distributionally Robust Optimization Framework for Secure Communications in Aerial-RIS Systems}

\author{Zhongming Feng,~\IEEEmembership{Graduate Student Member,~IEEE}, Qiling Gao*,~\IEEEmembership{Member,~IEEE}, Zeping Sui,~\IEEEmembership{Member,~IEEE}, \\ Yun Lin,~\IEEEmembership{Senior Member,~IEEE}, and Michail Matthaiou,~\IEEEmembership{Fellow,~IEEE}
\thanks{Zhongming Feng, Qiling Gao and Yun Lin are with the College of Information and Communication Engineering, Harbin Engineering University, Harbin 150000, China (e-mail: fzm98@hrbeu.edu.cn; qilinggao@outlook.com; linyun@hrbeu.edu.cn).}
\thanks{Zeping Sui is with the School of Computer Science and Electronics Engineering, University of Essex, Colchester CO4 3SQ, U.K. (e-mail: zepingsui@outlook.com).}
\thanks{Michail Matthaiou is with the Centre for Wireless Innovation (CWI), Queen’s University Belfast, BT3 9DT Belfast, U.K.  (e-mail: m.matthaiou@qub.ac.uk).}
\thanks{\emph{Corresponding author: Qiling Gao.}}

}



\maketitle

\begin{abstract}
This letter proposes a two-stage distributionally robust optimization (DRO) framework for secure deployment and beamforming in an aerial reconfigurable intelligent surface (A-RIS) assisted millimeter-wave system. To account for multi-timescale uncertainties arising from user mobility, imperfect channel state information (CSI), and hardware impairments, our approach decouples the long-term unmanned aerial vehicle (UAV) placement from the per-slot beamforming design. By employing the conditional value-at-risk (CVaR) as a distribution-free risk metric, a low-complexity algorithm is developed, which combines a surrogate model for efficient deployment with an alternating optimization (AO) scheme for robust real-time beamforming. Simulation results validate that the proposed DRO-CVaR framework significantly enhances the tail-end secrecy spectral efficiency and maintains a lower outage probability compared to benchmark schemes, especially under severe uncertainty conditions.
\end{abstract}


\section{Introduction}
\IEEEPARstart{A}{s} sixth-generation (6G) communication networks evolve, the millimeter-wave (mmWave) band is becoming a key technology thanks to its abundant spectral resources \cite{ref1}. However, signals in this band are susceptible to severe blockages by buildings in urban environments and also experience significant path loss. A reconfigurable intelligent surface (RIS) can mitigate these phenomena by intelligently manipulating the reflected signals to create virtual links, however, the performance of a ground-fixed RIS is limited by its inability to simultaneously maintain high-quality links with both the base station (BS) and users \cite{ref2}. In contrast, unmanned aerial vehicles (UAVs) offer flexible deployment and enable the rapid establishment of aerial line-of-sight (LoS) links. Consequently, deploying a RIS on a UAV—forming an aerial RIS (A-RIS)—facilitates dynamic adjustment of its spatial position and orientation, thereby mitigating mmWave blockages and enhancing both network coverage and physical-layer security (PLS) performance \cite{ref3}.

However, the practical deployment of an A-RIS is constrained by multi-source and multi-timescale uncertainties with unknown prior distributions \cite{ref4, ref5}. These uncertainties include channel state information (CSI) estimation errors from user mobility and estimation latency \cite{ref6}, RIS phase deviations due to UAV attitude jitter \cite{ref7}, phase-dependent amplitude (PDA) distortion in RIS elements \cite{ref8}, and distortion noise from transceiver hardware \cite{ref9}. Unfortunately, conventional approaches prove inadequate. Worst-case optimization is often overly conservative, while stochastic optimization demands precise distributional priors that are typically unavailable in practice \cite{ref10}. Furthermore, reinforcement learning (RL), despite its adaptability, is hampered by high computational overhead and lacks the performance guarantees essential for secure communications \cite{ref11}.

To address these fundamental, this letter proposes a two-stage distributionally robust optimization (DRO) framework \cite{ref12}. This framework decouples the long-term deployment location selection from the per-slot robust beamforming design. The main contributions are as follows:

1) For long-term secure communications in a static A-RIS system, a novel two-stage DRO framework is proposed. This framework  integrates Ricean fading, multifaceted hardware impairments, and spatially consistent user mobility. It employs conditional value-at-risk (CVaR) as the risk measure and constructs an ambiguity set through a projection-and-sampling method based on the Wasserstein ball. This approach allows for the explicit constraint of tail performance without relying on any specific distributional assumptions.

2) To solve the proposed beamforming design problem, we devise an efficient two-stage algorithm. Specifically, the algorithm first utilizes a lightweight surrogate model to significantly compress the search space. Subsequently, the fine-tuning stage integrates robust beamforming with adversarial sampling and Wirtinger phase updates through an alternating optimization (AO) algorithm. This method effectively handles long-term uncertainties arising from mobility, platform jitter, and hardware distortions, and consistently improves the low-quantile secrecy spectral efficiency (SSE) and tail security performance.

\section{System Model}
\begin{figure}[!t]
\centering
\includegraphics[width=2.3 in]{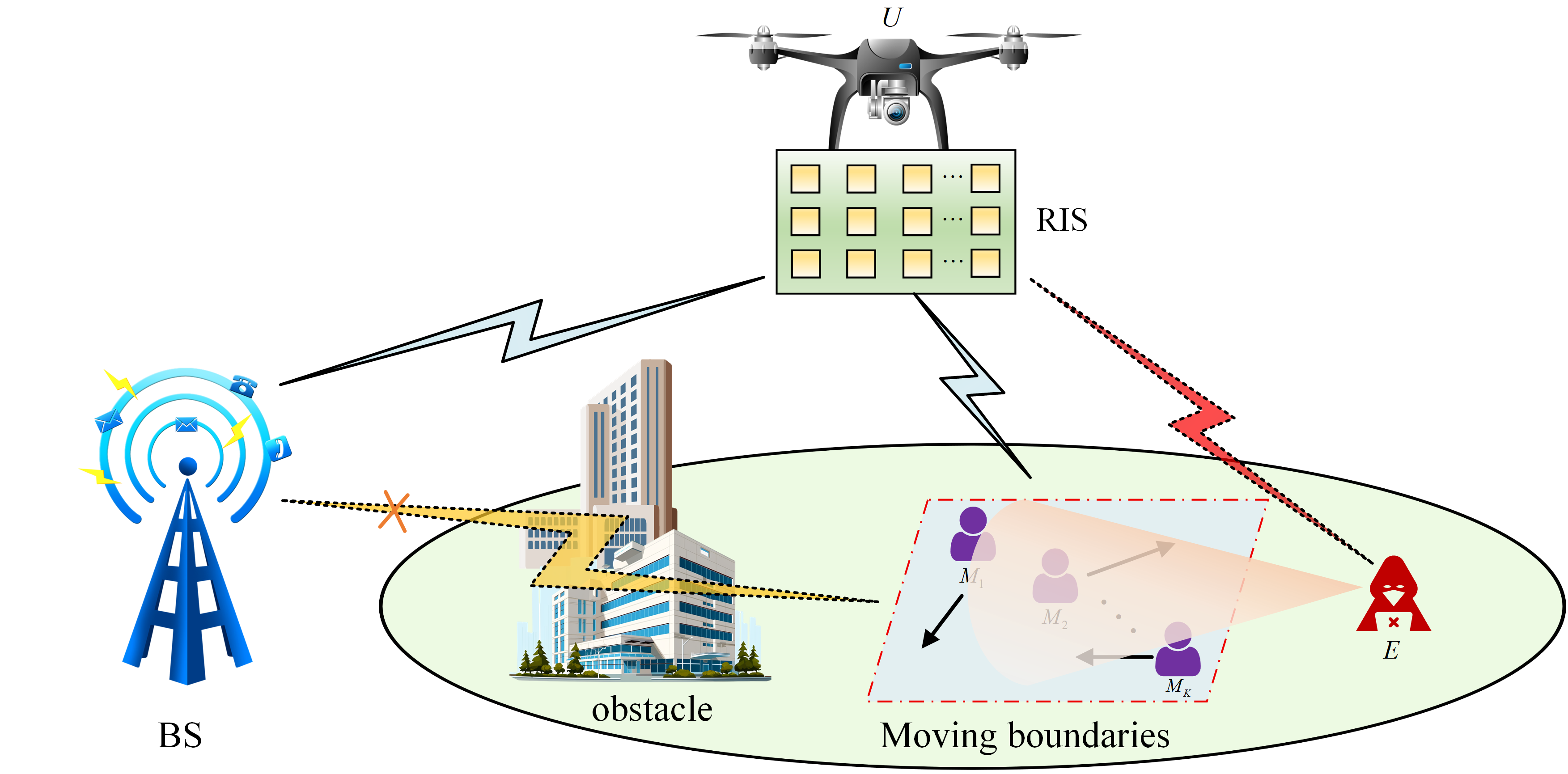}
\caption{Secure communication scenario with an A-RIS.}
\label{fig_1}
\vspace{-6pt}
\end{figure}

As shown in Figure~\ref{fig_1}, we consider a time-slotted secure mmWave downlink aided by an A-RIS. A ground BS equipped with an $N$-element uniform linear array (ULA) serves $K$ single-antenna mobile users $\{M_k\}_{k=1}^{K}$ in the presence of a single-antenna eavesdropper $E$. We assume that the direct links between the BS and all ground terminals are obstructed by obstacles. A UAV $U$ carries a RIS with $N_R=N_xN_y$ reflecting elements located at $\boldsymbol r_U\in\mathbb{R}^3$, which establishes the necessary cascaded reflective paths. The RIS reflection pattern at slot $t$ is $\boldsymbol v(t)=[v_1(t),\ldots,v_{N_R}(t)]^{\mathrm T}\!\in\mathbb{C}^{N_R}$ with $|v_n(t)|\le 1$, and $\boldsymbol\Phi(t)=\mathrm{diag}[\boldsymbol v(t)]$. The channels of the BS-RIS, RIS-$M_k$, RIS-$E$ links are denoted by $\mathbf H_{\mathrm{BR}}(t)\in\mathbb{C}^{N_R\times N}$, $\boldsymbol h_{\mathrm{RM}_k}(t)\in\mathbb{C}^{N_R\times 1}$ and $\boldsymbol h_{\mathrm{RE}}(t)\in\mathbb{C}^{N_R\times 1}$, respectively.
\vspace{-10pt}

\subsection{Channel and Signal Model}
We adopt a narrowband block-fading mmWave air-to-ground (A2G) model \cite{ref7}, assuming the channel is static within each slot of duration $T_s=10\,\mathrm{ms}$. With a carrier frequency $f_c=28\,\mathrm{GHz}$ and mean user speed $v=1.0\,\mathrm{m/s}$, the channel coherence time $T_c \approx c/(v f_c) \approx 10.7\,\mathrm{ms}$. Each link is Ricean with a rank-one LoS term induced by the BS-ULA and RIS-UPA steering vectors. The BS–RIS, RIS–$M_k$, and RIS–$E$ channels are given by\label{eq:rician_all}
\begin{equation} \label{eq:rician_all}
\mathbf H_{\mathrm{BR}}(t)
= \sqrt{L_{\mathrm{BR}}(t)}\Big(\alpha_{\mathrm{BR}}\,\mathbf H_{\mathrm{BR}}^{\mathrm{LoS}}(t)
+\bar\alpha_{\mathrm{BR}}\,\mathbf H_{\mathrm{BR}}^{\mathrm{NLoS}}(t)\Big),
\end{equation}
\begin{equation} \label{eq:rician_RMk}
\mathbf h_{\mathrm{RM}_k}(t)
= \sqrt{L_{\mathrm{RM}_k}(t)}\Big(\alpha_{\mathrm{RM}}\,\mathbf a_R(\varphi_k,\theta_k)
+\bar\alpha_{\mathrm{RM}}\,\mathbf z_k(t)\Big), 
\end{equation}
\begin{equation} \label{eq:rician_RE}
\mathbf h_{\mathrm{RE}}(t)
= \sqrt{L_{\mathrm{RE}}(t)}\Big(\alpha_{\mathrm{RE}}\,\mathbf a_R(\varphi_E,\theta_E)
+\bar\alpha_{\mathrm{RE}}\,\mathbf z_E(t)\Big), 
\end{equation}
where $L_x(t)$ encapsulates the large-scale effects, while $\alpha_x=\sqrt{\kappa_x/(\kappa_x+1)}$ and $\bar \alpha_x=1/\sqrt{\kappa_x+1}$ partition the power between the LoS and NLoS components, governed by the Ricean factor $\kappa_x$ where $x \in \{\mathrm{BR}, \mathrm{RM}, \mathrm{RE}\}$. The NLoS component $\mathbf H_{\mathrm{BR}}^{\mathrm{NLoS}}(t)$ contains entries distributed as $\mathcal{CN}(0,1)$, while the vectors $\mathbf z_k(t), \mathbf z_E(t) \in \mathbb{C}^{N_R \times 1}$ are distributed as $\mathcal{CN}(\mathbf{0}, \mathbf{I}_{N_R})$. The LoS component $\mathbf H_{\mathrm{BR}}^{\mathrm{LoS}}(t)=\mathbf a_R(\varphi_R,\theta_R)\,\mathbf a_B^{H}(\varphi_B)$ is a rank-one matrix formed by the unit-norm array response vectors.

The BS transmits a linearly precoded signal $\mathbf{x}(t) = \sum_{k=1}^K \mathbf{w}_k(t) s_k(t)$, where $\mathbf{w}_k(t) \in \mathbb{C}^{N \times 1}$ is the precoder for the unit-power symbol $s_k(t)$ of user $M_k$, subject to a total power constraint $\sum_{k=1}^K \|\mathbf{w}_k(t)\|^2 \le P_{\max}$. The RIS reflects this signal via the matrix $\boldsymbol{\Phi}(t) $, forming the idealized end-to-end cascaded channel $\mathbf{g}_{j}^{H}(t) = \mathbf{h}_{\mathrm{R}j}^H(t) \boldsymbol{\Phi}(t) \mathbf{H}_{\mathrm{BR}}(t)$ for $j \in \{M_k, E\}$. The received signals at user $M_k$ and the eavesdropper $E$ are respectively given by\label{eq:rx_signals}
\begin{equation}\label{eq:rx_signals}
\begin{aligned}
y_{M_k}(t) &=
  \underbrace{\mathbf g_{M_k}^{H}(t)\mathbf w_k(t)s_k(t)}_{\text{Legitimate Signal}}
  +\underbrace{\sum_{j\neq k}\mathbf g_{M_k}^{H}(t)\mathbf w_j(t)s_j(t)}_{\text{Multi-user Interference}}
  + n_k(t),\\[2pt]
y_{E}(t) &=
  \underbrace{\sum_{j=1}^{K}\mathbf g_{E}^{H}(t)\mathbf w_j(t)s_j(t)}_{\text{Intercepted Signal}}
  + n_E(t),
\end{aligned}
\end{equation}
where $n_k(t), n_E(t)\sim\mathcal{CN}(0, \sigma^2)$ are the additive noise samples. The corresponding instantaneous SINRs, $\gamma_{M_k}(t)$ and $\gamma_{E,k}(t)$, are derived accordingly as
\begin{equation}\label{eq:sinr}
\begin{aligned}
\gamma_{M_k}(t)
&=\frac{\left|\mathbf g_{M_k}^{H}(t)\mathbf w_k(t)\right|^2}
{\sum_{j\neq k}\left|\mathbf g_{M_k}^{H}(t)\mathbf w_j(t)\right|^2+\sigma^2},\\
\gamma_{E,k}(t)
&=\frac{\left|\mathbf g_{E}^{H}(t)\mathbf w_k(t)\right|^2}
{\sum_{j\neq k}\left|\mathbf g_{E}^{H}(t)\mathbf w_j(t)\right|^2+\sigma^2}.
\end{aligned}
\end{equation}

The SINR expressions in \eqref{eq:sinr} establish an idealized performance benchmark. However, in a practical deployment, this benchmark is fundamentally challenged by multifaceted hardware impairments and channel uncertainties, which we now proceed to model.

\subsection{Long-Term Uncertainties and Hardware Impairments}
To bridge the gap between the ideal and the actual scenarios, we now introduce a more realistic model that incorporates the above imperfections by first detailing their physical origins and then establishing a unified error model to capture their collective impact.

\paragraph{User Mobility and RIS Hardware Impairments}
Two primary physical phenomena corrupt the ideal channel model. The first is user mobility, where the positions of ground users $\{\mathbf{r}_{M_k}(t)\}$ evolve according to a Gauss-Markov model, introducing slow-varying changes to the channel's geometric parameters. The second is multifaceted RIS hardware impairments. In a practical environment, the actual reflection coefficient $\tilde{v}_n(t)$ of the $n$-th RIS element, deviates from its intended value, which can be expressed as
\begin{equation} \label{eq:ris_impairments}
    \tilde{v}_n(t) = \beta(\phi_n(t)) e^{j\phi_n(t)}, \text{with} \quad \phi_n(t) = \hat{\phi}_n(t) + \Delta\phi_n(t).
\end{equation}

This model encapsulates a cascade of hardware limitations: 1) Discrete phase quantization, where the intended phase $\hat{\phi}_n(t)$ is selected from a finite $B$-bit codebook $\mathcal{F}$; 2) Phase noise, where the error $\Delta\phi_n(t)$ follows a von Mises distribution, $\Delta\phi_n(t) \sim \mathrm{VM}(0, \kappa)$ \cite{ref13}; and 3) Phase-dependent amplitude, where the reflection amplitude $\beta(\cdot)$ varies non-linearly with the actual phase $\phi_n(t)$. These impairments collectively form the true unknown reflection vector $\tilde{\boldsymbol{v}}(t)$, which in turn defines the true physical cascaded channel, yielding:
\begin{equation} \label{eq:true_physical_channel}
    \mathbf{g}_{j}^H(t) = \mathbf{h}_{\mathrm{R}j}^H(t) \mathrm{diag}(\tilde{\boldsymbol{v}}(t)) \mathbf{H}_{\mathrm{BR}}(t).
\end{equation}

\paragraph{Unified Model for Imperfect CSI}
Due to the aforementioned physical phenomena, compounded by inherent channel estimation and feedback latency, the BS cannot perfectly know the true channel $\mathbf{g}_{j}(t)$. Instead, it only has access to an imperfect channel estimate, denoted by $\widehat{\mathbf{g}}_{j}(t)$. To capture the total discrepancy, we adopt the widely used additive error model:
\begin{equation} \label{eq:csi_error}
    \mathbf{g}_{j}(t)=\widehat{\mathbf{g}}_{j}(t)+\Delta\mathbf{g}_{j}(t),
\end{equation}
where the error vector $\Delta\mathbf g_j(t)\triangleq \mathbf g_j(t)-\widehat{\mathbf g}_j(t)$ represents the discrepancy between the true channel \eqref{eq:true_physical_channel} and its BS-side estimate. Since the probability law of this error is intractable in practice, we adopt a distribution-free robust model, constraining the error to lie within the Euclidean ball
\begin{equation} \label{eq:uncertainty_set}
    \mathcal{G}_j = \{ \Delta\mathbf{g}_{j}(t) \mid \|\Delta\mathbf{g}_{j}(t)\|_2^2 \le \sigma_{e,j}^2 \}, \quad \forall j \in \{M_k, E\},
\end{equation}
where $\sigma_{e,j}^2$ is specified in Table~\ref{tab:sim_params}. This bounded-error model underpins the worst-case formulation in \eqref{eq:prob_main}. The precoder is designed based on the estimate $\widehat{\mathbf{g}}_{j}(t)$, though the system actual performance is dictated by the true channel $\mathbf{g}_{j}(t)$. Consequently, the actual instantaneous SINRs including this total uncertainty are given by

\begin{equation}\label{eq:actual_sinr}
\begin{aligned}
\tilde{\gamma}_{M_k}(t)
&=\frac{\left|(\widehat{\mathbf g}_{M_k}(t)+\Delta\mathbf{g}_{M_k}(t))^{H}\mathbf w_k(t)\right|^2}
{\sum_{j\neq k}\left|(\widehat{\mathbf g}_{M_k}(t)+\Delta\mathbf{g}_{M_k}(t))^{H}\mathbf w_j(t)\right|^2+\sigma^2},\\
\tilde{\gamma}_{E,k}(t)
&=\frac{\left|(\widehat{\mathbf g}_{E}(t)+\Delta\mathbf{g}_{E}(t))^{H}\mathbf w_k(t)\right|^2}
{\sum_{j\neq k}\left|(\widehat{\mathbf g}_{E}(t)+\Delta\mathbf{g}_{E}(t))^{H}\mathbf w_j(t)\right|^2+\sigma^2}.
\end{aligned}
\end{equation}

The significant deviation of these actual SINRs from their idealized counterparts motivates our robust design.

\subsection{Secrecy Metric and Problem Statement}
The system performance is evaluated by the actual instantaneous sum SSE, which is a function of the total channel uncertainty, yielding:
\begin{equation} \label{eq:secrecy_rate}
    \tilde{R}_{\mathrm{sec}}(t) = \sum_{k=1}^K \Big[ \log_2(1+\tilde{\gamma}_{M_k}(t)) - \log_2(1+\tilde{\gamma}_{E,k}(t)) \Big]^+,
\end{equation}
where $[\cdot]^+ \triangleq \max(0, \cdot)$ and the SINR terms are defined in \eqref{eq:actual_sinr}. Given the CSI uncertainty model established by \eqref{eq:csi_error} and \eqref{eq:uncertainty_set}, the BS lacks access to the true $\mathbf{g}_{j}(t)$. We therefore adopt a robust design based on the worst-case performance.

The overarching goal is to find an optimal static deployment location $\boldsymbol r_U$ for the A-RIS, and a corresponding set of robust beamforming strategies over a communication horizon $T$ to maximize the long-term secrecy performance. This joint optimization problem can be formulated as
\begin{IEEEeqnarray}{rCl}\label{eq:prob_main}
\underset{\{ \boldsymbol r_U,\{\mathbf W(t),\boldsymbol{v}(t)\}_{t=1}^T\}}{\max} &\quad&
\frac{1}{T}\sum_{t=1}^{T}\min_{\Delta \mathbf g_j(t) \in \mathcal G_j, \forall j}\,\tilde R_{\mathrm{sec}}(t)
\IEEEyesnumber\label{eq:prob_obj}\\[-0.4ex]
\text{s.t.} && \|\mathbf W(t)\|_{F}^{2} \le P_{\max},\ \forall t,\ \IEEEeqnarraynumspace
\IEEEyessubnumber\label{eq:prob_power_const}\\[-0.25ex]
&& v_n(t)=e^{j\hat\phi_n(t)},\ \hat\phi_n(t)\in\mathcal F,\ \forall n,t,\ \IEEEeqnarraynumspace
\IEEEyessubnumber\label{eq:prob_ris_const}\\[-0.25ex]
&& \boldsymbol r_U \in \mathcal P,\ \IEEEeqnarraynumspace
\IEEEyessubnumber\label{eq:prob_uav_pos_const}
\end{IEEEeqnarray}
where $\mathbf{W}(t) \triangleq [\mathbf{w}_1(t), \dots, \mathbf{w}_K(t)] \in \mathbb{C}^{N \times K}$, while $\mathcal{P}$ is the feasible deployment area for the UAV. The constraints define the physical limits of the system: \eqref{eq:prob_power_const} constraints the maximum transmit power at the BS; \eqref{eq:prob_ris_const} models the hardware limitations of the RIS, where each element must select a phase from a discrete codebook $\mathcal{F}$ determined by the hardware quantization bits; and \eqref{eq:prob_uav_pos_const} restricts the UAV's static position to a predefined geographical region. Problem \eqref{eq:prob_main} is a non-convex problem with coupled multi-timescale variables and discrete constraints, which is intractable to solve directly. Therefore, we propose to tackle it using a two-stage optimization framework.

\section{Proposed Two-Stage DRO-CVaR Framework}
To tackle the multi-timescale optimization challenge in \eqref{eq:prob_main}, we employ a two-stage DRO framework decoupling the long-term offline deployment optimization from the per-slot real-time beamforming. The complexity of the real-time algorithm is analyzed in Sec.~\ref{sec:complexity} to verify its feasibility within the channel coherence time $T_c$. In this framework, we frequently solve a per-slot subproblem, for which the constraints \eqref{eq:prob_power_const} and \eqref{eq:prob_ris_const} are respectively written as

\begin{equation} \label{eq:per_slot_constraints}
\begin{aligned}
\|\mathbf W\|_{F}^{2} &\le P_{\max},  \\
v_n &= e^{j\hat\phi_n},\ \hat\phi_n\in\mathcal F,\ \forall n.
\end{aligned}
\end{equation}

\paragraph*{Stage 1: Long-Term Deployment via Surrogate-Assisted CVaR Optimization}
Direct optimization of the static location $\boldsymbol r_U$ is prohibitive, thus we adopt a two-step screening. First, a coarse screening on $\mathcal P$ finds a subset $\mathcal P_{\text{fine}}$ via a risk-aware surrogate $\mathcal S(\boldsymbol r_U)$. To compute it, we probe each $\boldsymbol r_U$ to find a warm-start RIS phase $\boldsymbol\Phi^{\rm tr}$ and best-probe SINR $\mathcal S_{\rm probe}(\boldsymbol r_U)$. We then collect $N_{\text{pre}}$ SSE samples $\{\tilde R_{\mathrm{sec}}(t)\}$ from \eqref{eq:secrecy_rate} using this fixed $\boldsymbol\Phi^{\rm tr}$ and maximum-ratio transmission (MRT) beams. We form the surrogate function $\mathcal S(\boldsymbol r_U)$ as
\begin{equation}\label{eq:surrogate_metric}
\mathcal S(\boldsymbol r_U)=(1-\omega)\,\mathrm{CVaR}_\alpha\!\big(\{\tilde R_{\mathrm{sec}}(t)\}_{t=1}^{N_{\text{pre}}}\big)+\omega\,\mathcal S_{\rm probe}(\boldsymbol r_U),
\end{equation}
where $\omega\!\in\![0,1]$. For $L$ samples $\{R_l\}_{l=1}^L$, the empirical worst-$\alpha$ tail $\mathrm{CVaR}$, with risk level $\alpha\in(0,1)$, is
\begin{equation}\label{eq:cvar_def}
\mathrm{CVaR}_{\alpha}(\{R_l\}_{l=1}^L)=\max_{\tau\in\mathbb R}\left\{\tau-\frac{1}{\alpha L}\sum_{l=1}^{L}[\tau-R_l]^+\right\},
\end{equation}
where $[\cdot]^+=\max\{\cdot,0\}$ and $L=N_{\text{pre}} = 48$ for \eqref{eq:surrogate_metric}. The subset is $\mathcal P_{\text{fine}}=\underset{\boldsymbol r_U\in\mathcal P}{\text{Top-}K_c}\{\mathcal S(\boldsymbol r_U)\}$, and a fine evaluation is performed on $\mathcal P_{\text{fine}}$. A few samples warm-start the robust AO routine, and we then run this routine over $N_{\text{fine}} = 200$ evaluation slots to select
\begin{equation}\label{eq:prob_stage1_fine}
\boldsymbol r_U^*=\arg\max_{\boldsymbol r_U\in\mathcal P_{\text{fine}}}\ \mathrm{CVaR}_\alpha\!\left(\left\{\tilde R_{\mathrm{sec},t}^{\star}(\boldsymbol r_U)\right\}_{t=1}^{N_{\text{fine}}}\right),
\end{equation}
where $\tilde R_{\mathrm{sec},t}^{\star}$ is the resulting SSE at slot $t$.

\paragraph*{Stage 2: DRO-based Per-Slot Beamforming}
With the UAV position fixed, this stage addresses the per-slot beamforming problem using our proposed DRO formulation. To tractably solve this problem, we employ a sample average approximation (SAA) approach, performing adversarial sampling from a Wasserstein ball-based ambiguity set to generate $N_s$ samples $\{\{\Delta\mathbf{g}_j^{(s)}\}_j\}_{s=1}^{N_s}$ \cite{ref13}. The problem is then formulated as
\begin{IEEEeqnarray}{rCl}
\label{eq:prob_cvar_epi} 
\underset{\{\mathbf{W}, \boldsymbol{v}, \tau, \{s_s\}\}}{\max} & \> & J(\mathbf{W}, \boldsymbol{v}, \tau, \{s_s\}) \triangleq \tau - \frac{1}{N_s(1-\alpha)}\sum_{s=1}^{N_s} s_s \IEEEeqnarraynumspace
\IEEEyesnumber \label{eq:obj_J} \\ 
\mathrm{s.t.} & & \tau - \tilde{R}_{\mathrm{sec}}(\mathbf{W}, \boldsymbol{v}; \{\Delta\mathbf{g}_j^{(s)}\}_j) \le s_s, \quad \forall s, \IEEEeqnarraynumspace
\IEEEyessubnumber \label{eq:const_epi} \\ 
& & s_s \ge 0, \ \forall s, \IEEEeqnarraynumspace
\IEEEyessubnumber \label{eq:const_other} \\
& & \eqref{eq:prob_power_const}, \eqref{eq:prob_ris_const}. \IEEEeqnarraynumspace
\IEEEyessubnumber \label{eq:const_ref}
\end{IEEEeqnarray}

The resulting optimization problem \eqref{eq:prob_cvar_epi} is solved via an efficient AO algorithm. First, for a fixed RIS vector $\boldsymbol{v}$, we solve for the precoder $\mathbf{W}$ using a threat-aware MRT scheme. Second, with $\mathbf{W}$ fixed, we update the RIS phases $\hat{\boldsymbol{\phi}}$ using a projected gradient step based on Wirtinger calculus, where the step size is determined via a backtracking line search. To ensure strict monotonicity, we apply a standard accept-if-improve rule after projection to $\mathcal{F}$. The analytical gradient is derived as
\begin{equation}
    \frac{\partial \tilde{R}_{\mathrm{sec}}}{\partial \hat{\phi}_n} = \Re\left\{ \frac{\partial \tilde{R}_{\mathrm{sec}}}{\partial v_n^*} \frac{\partial v_n^*}{\partial \hat{\phi}_n} \right\}, \text{ with } \frac{\partial v_n^*}{\partial \hat{\phi}_n} = -j e^{-j\hat{\phi}_n}.
\end{equation}

Finally, after updating $\mathbf{W}$ and $\boldsymbol{v}$, the epigraph variables $\{\tau, \{s_s\}\}$ are updated optimally in closed form to maximize \eqref{eq:obj_J} subject to \eqref{eq:const_epi}.

\paragraph*{Convergence Analysis}
The convergence of the AO algorithm, which solves the stage-2 subproblem \eqref{eq:prob_cvar_epi}, is analyzed as follows. The proposed algorithm is a block coordinate ascent method. Let $\Theta^{(i)} = \{\mathbf{W}^{(i)}, \boldsymbol{v}^{(i)}, \tau^{(i)}, \{s_s\}^{(i)}\}$ denote the collection of variables at the beginning of iteration $i$, and let $J(\Theta)$ be the objective function in \eqref{eq:obj_J}. In each iteration, the blocks of variables are updated sequentially. The first step updates the precoder from $\mathbf{W}^{(i)}$ to $\mathbf{W}^{(i+1)}$ by applying the threat-aware MRT scheme. As this step is designed to improve the objective, we have
\begin{multline} \label{eq:conv_step1}
    J(\mathbf{W}^{(i+1)}, \boldsymbol{v}^{(i)}, \tau^{(i)}, \{s_s\}^{(i)}) \\
    \ge J(\mathbf{W}^{(i)}, \boldsymbol{v}^{(i)}, \tau^{(i)}, \{s_s\}^{(i)}).
\end{multline}

Subsequently, for a fixed precoder $\mathbf{W}^{(i+1)}$, the RIS vector is updated to $\boldsymbol{v}^{(i+1)}$ using a projected gradient step. The use of a backtracking line search and an accept-if-improve rule for the projection onto the discrete set $\mathcal{F}$ strictly guarantees a non-decreasing objective value, which leads to
\begin{multline} \label{eq:conv_step2}
    J(\mathbf{W}^{(i+1)}, \boldsymbol{v}^{(i+1)}, \tau^{(i)}, \{s_s\}^{(i)}) \\
    \ge J(\mathbf{W}^{(i+1)}, \boldsymbol{v}^{(i)}, \tau^{(i)}, \{s_s\}^{(i)}).
\end{multline}

Finally, the epigraph variables are updated to $\{\tau^{(i+1)}, \{s_s\}^{(i+1)}\}$ using their closed-form solution, which by definition maximizes the objective with respect to this block. Thus, we have
\begin{multline} \label{eq:conv_step3}
    J(\mathbf{W}^{(i+1)}, \boldsymbol{v}^{(i+1)}, \tau^{(i+1)}, \{s_s\}^{(i+1)}) \\
    \ge J(\mathbf{W}^{(i+1)}, \boldsymbol{v}^{(i+1)}, \tau^{(i)}, \{s_s\}^{(i)}).
\end{multline}

Combining these inequalities yields $J(\Theta^{(i+1)}) \ge J(\Theta^{(i)})$, which proves that the sequence of objective values $\{J(\Theta^{(i)})\}$ is monotonically non-decreasing. Given that the objective function $J$ is bounded above due to the transmit power constraint in $\|\mathbf W\|_{F}^{2} \le P_{\max}$, the sequence $\{J(\Theta^{(i)})\}$ is guaranteed to converge by the Monotone Convergence Theorem.

\paragraph*{Complexity Analysis}
\label{sec:complexity}
Execution of the per-slot robust AO algorithm is constrained by $T_c \approx 10.7\,\mathrm{ms}$. The complexity per iteration is dominated by the $\mathcal{O}(N_s N_R N K (K+E))$ Wirtinger gradient computation across $N_s$ samples for the RIS update, targeting the CVaR objective \eqref{eq:obj_J}. Thus, the total per-slot complexity scales as $\mathcal{O}(I_{\text{AO}} N_s N_R N K (K+E))$. The algorithm converges rapidly with a small number of iterations $I_{\text{AO}}$, requires a modest sample size $N_s$, and features a highly parallelizable dominant gradient computation, thus enabling completion within the $T_c$ constraint.

\section{Simulation Results And Discussions}
\label{sec:sim}
In this section, we provide numerical results of our proposed two-stage DRO framework. We consider a mmWave system with a carrier frequency of $f_c = 28\,\mathrm{GHz}$ and a bandwidth of $B = 100\,\mathrm{MHz}$. The BS is located at $\mathbf{r}_{BS} = [0, 0, 30]^\mathrm{T}$\,m. Users are initially distributed in a blockage sector relative to the BS, within an angular sector of [$35^\circ, 85^\circ$] and at a distance of 60 -- 100\,m. User mobility follows a Gauss-Markov model with a mean speed of 1.0 m/s. The eavesdropper is randomly placed near the users with a 20 -- 40\,m offset. The maximum BS transmit power is set to $P_{\max} = 2.0\,\mathrm{W}$, with a per-antenna limit of $0.1\,\mathrm{W}$. The total noise power $\sigma^2$ is $-84\,\mathrm{dBm}$, based on a thermal noise density of $N_0 = -174\,\mathrm{dBm/Hz}$ and a receiver noise figure of $N_F = 10\,\mathrm{dB}$. For the channel model, the Ricean K-factors are set to $\kappa_{\mathrm{BR}}=10\,\mathrm{dB}$, $\kappa_{\mathrm{RU}}=3\,\mathrm{dB}$, and $\kappa_{\mathrm{RE}}=5\,\mathrm{dB}$, with corresponding path-loss exponents $\alpha_{\mathrm{BR}}=2.5$, $\alpha_{\mathrm{RU}}=3.5$, and $\alpha_{\mathrm{RE}}=3.2$. For our proposed DRO-CVaR framework, the number of samples for SAA is $N_s=10$, the Wasserstein radius for the ambiguity set is $\epsilon=0.03$ and the number of candidates selected in the Stage-1 coarse search is $K_c=12$. The remaining key simulation parameters are summarized in Table~\ref{tab:sim_params}.

\begin{table}[!t]
\centering
\caption{Key Simulation Parameters}
\label{tab:sim_params}
\renewcommand{\arraystretch}{1.1}
\begin{tabular}{l l}
\hline
\textbf{Parameter} & \textbf{Value} \\
\hline
\multicolumn{2}{c}{\textbf{System Layout and Antennas}} \\
\hline
Number of BS Antennas, $N$ & 32 \\
Number of RIS Elements, $N_R$ & 64 (8 $\times$ 8) \\
Number of Users, $K$ & 2 \\
Number of Eavesdroppers, $E$ & 1 \\
UAV Altitude Range & [60, 150] m \\
User Mobility Area & [-75, 75]~m $\times$ [-75, 75]~m \\
\hline
\multicolumn{2}{c}{\textbf{Hardware Impairment and Uncertainty}} \\
\hline
RIS Phase Quantization Bits, $B$ & 2 \\
RIS Phase Jitter Concentration, $\kappa$ & 5.0 \\
Additional Phase Noise Std. Dev. & \SI{0.3}{rad} \\
Amplitude Error Std. Dev. & 0.15 \\
RIS Amplitude Range, $[\beta_{\min}, \beta_{\max}]$ & [0.4, 0.8] \\
CSI Error Variance, User Link ($\sigma_{e,M_k}^2$) & 0.10 \\
CSI Error Variance, Eve Link ($\sigma_{e,E}^2$) & 0.12 \\
CSI Error Variance, BS-RIS Link ($\sigma_{e,BR}^2$) & 0.05 \\
CSI Correlation Factor, $\rho$ & 0.75 \\
BS Error Vector Magnitude (EVM) & 8\% \\
User Error Vector Magnitude (EVM) & 10\% \\
\hline
\end{tabular}
\end{table}

\begin{figure}[!t]
\centering
\includegraphics[width=0.7\columnwidth]{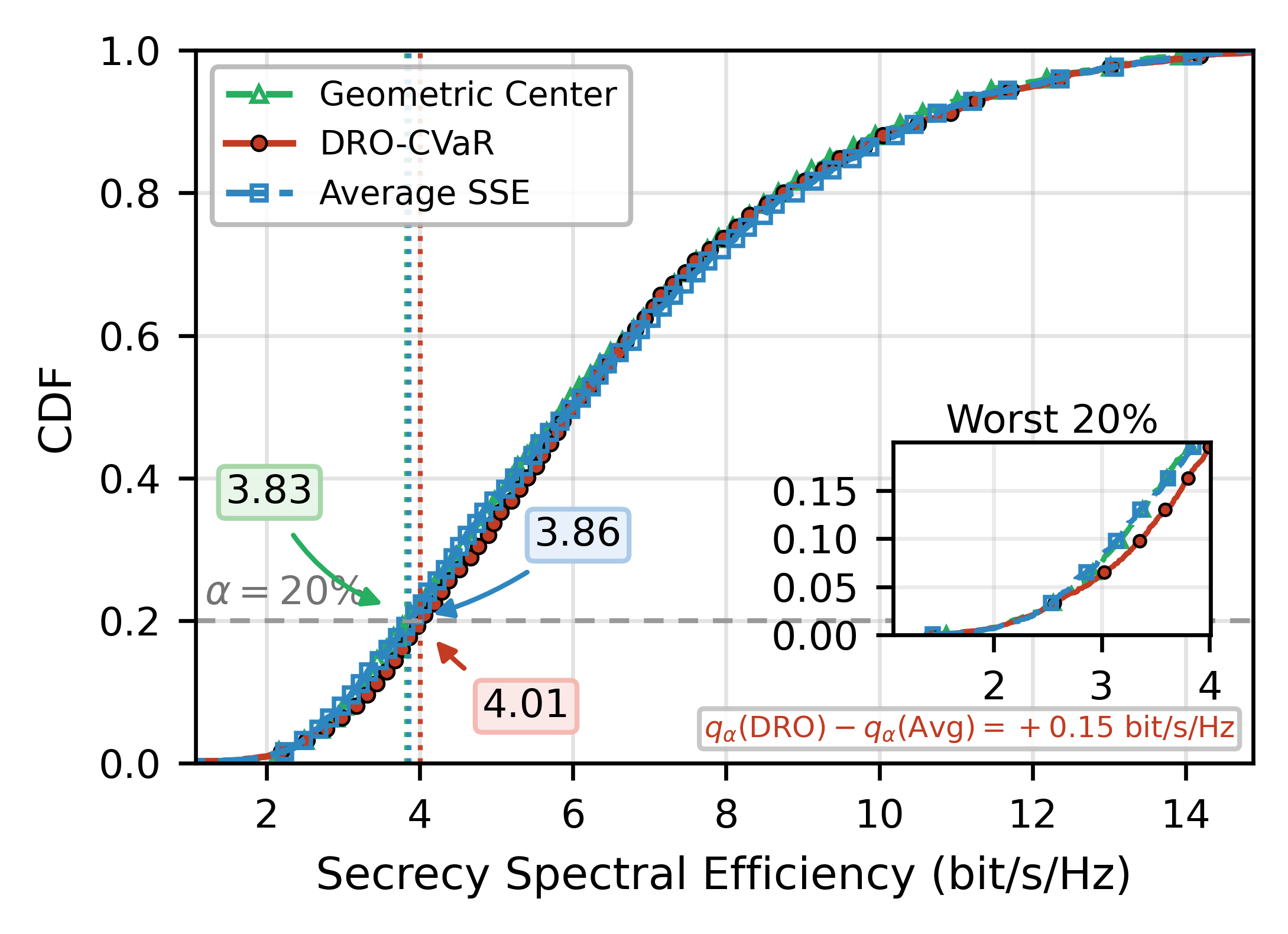}
\caption{CDF of long-term SSE for different A-RIS deployment strategies.}
\label{fig:deployment_cdf}
\vspace{-18pt}
\end{figure}

Figure~\ref{fig:deployment_cdf} compares the cumulative distribution function (CDF) of the long-term SSE for the proposed DRO-CVaR framework against two benchmark deployment strategies, evaluated across 50 candidate UAV positions. Results are averaged over three independent trials to ensure statistical validity, with the final performance shown after a 200-slot deployment evaluation per candidate followed by a 1000-slot evaluation phase. Targeting the 20th percentile performance via the CVaR metric with risk level $\alpha=0.2$, DRO-CVaR achieves 4.10 bit/s/Hz. This yields gains of 0.24 bit/s/Hz and 0.41 bit/s/Hz over the average SSE and geometric center methods \cite{ref14}, which achieve 3.86 bit/s/Hz and 3.69 bit/s/Hz, respectively. This result directly validates that the deployment location selected by optimizing the CVaR metric yields superior robustness.

\begin{figure}[!t]
\centering
\includegraphics[width=0.85\columnwidth]{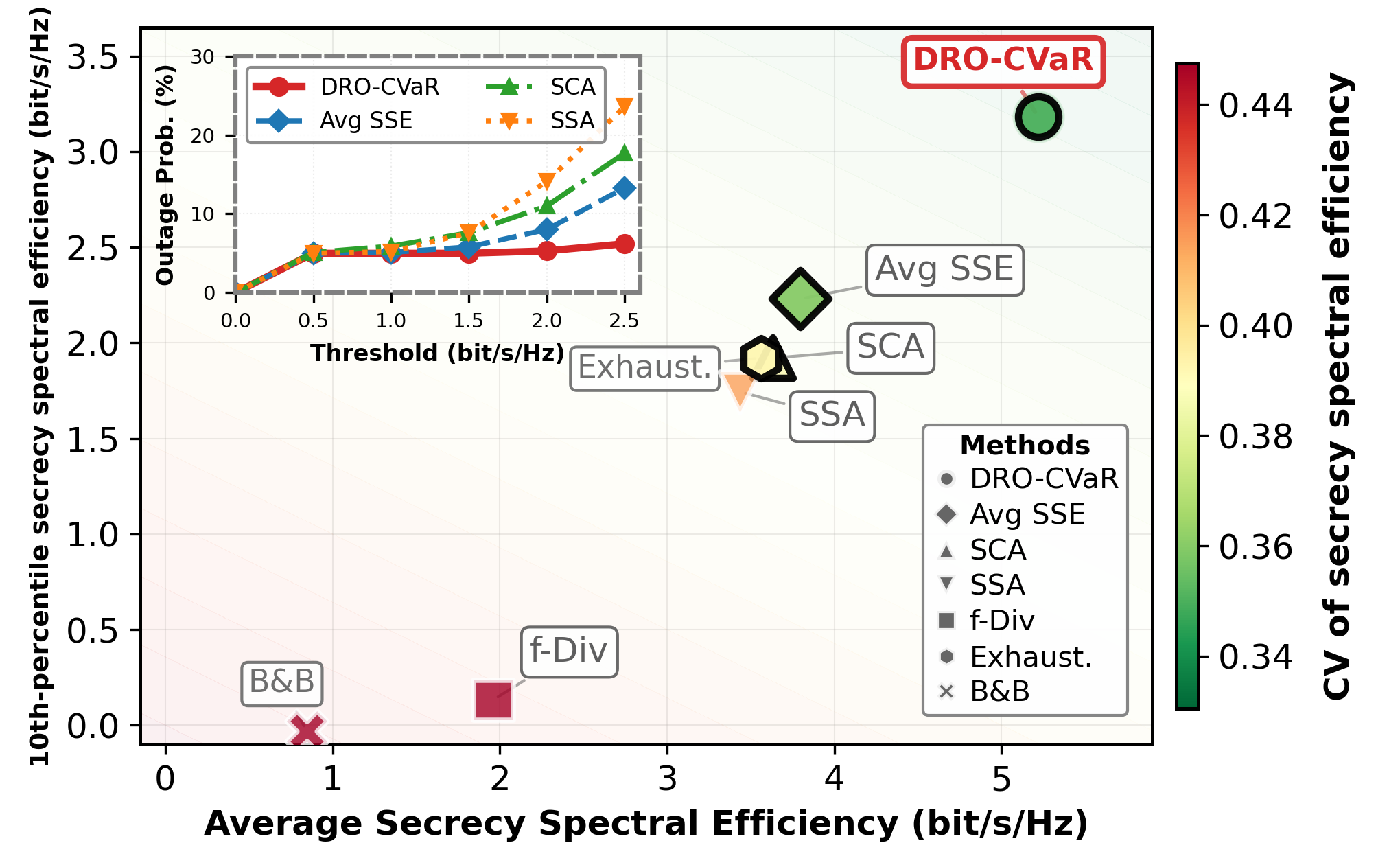}
\caption{Robustness and outage probability comparison of different schemes.}
\label{fig:robustness_tradeoff}
\end{figure}
Figure~\ref{fig:robustness_tradeoff} presents the trade-off between average SSE and robustness for the proposed DRO-CVaR algorithm against several benchmarks, evaluated over three independent 1500-slot trials. The proposed algorithm achieves a 10th-percentile SSE of 3.18 bit/s/Hz, which is 65.6\% higher than the second-best SCA benchmark. This result validates the significant advantage of the proposed method in robust tail optimization, while a low coefficient of variation (CV) of 0.35 further confirms its high stability. Furthermore, the inset indicates that our method yields the lowest outage probability, which is only 6.2\% at a 2.5 bit/s/Hz threshold, compared to 23.6\% for the SAA scheme.

\begin{figure}[!t]
\centering
\includegraphics[width=0.85\columnwidth]{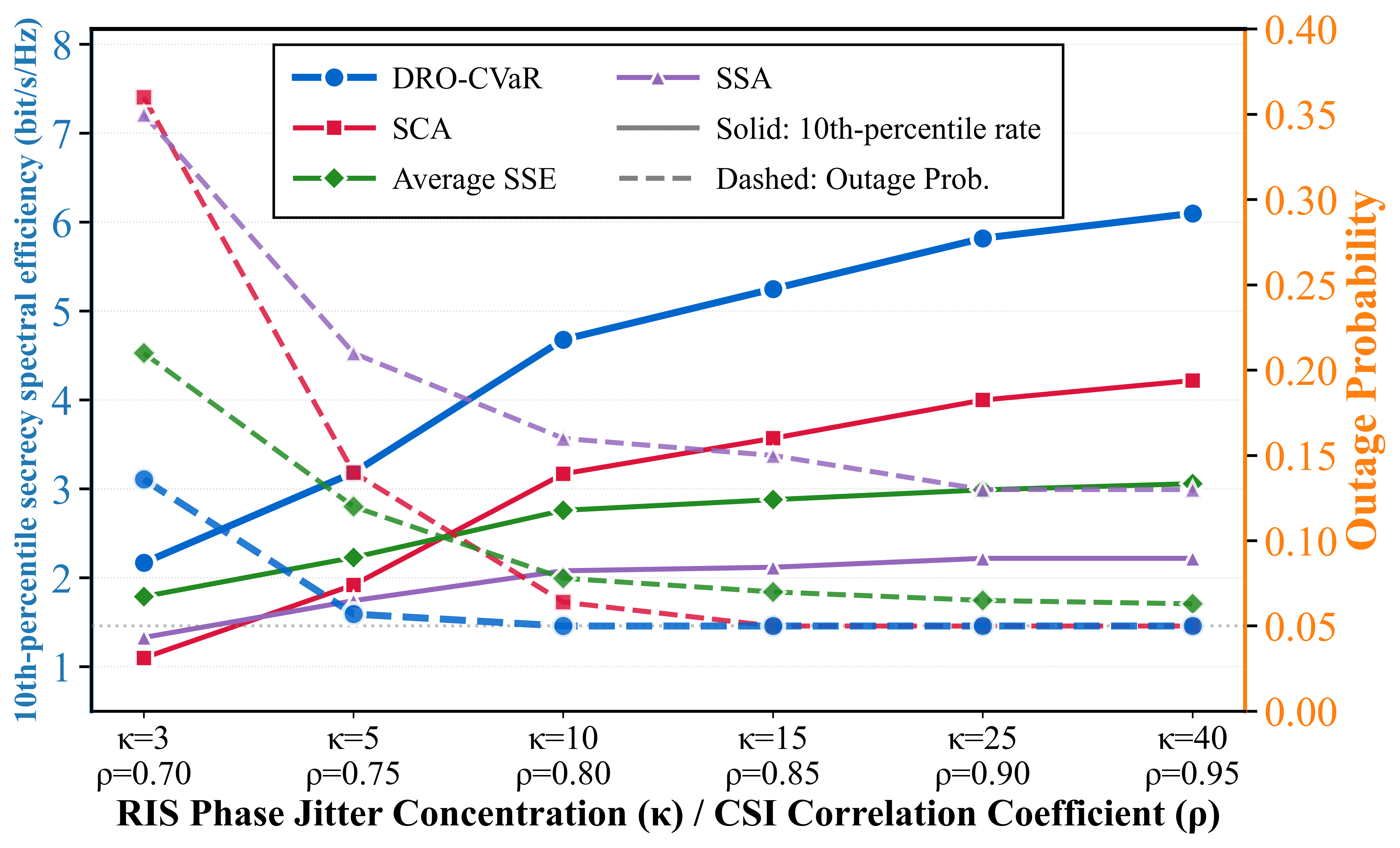}
\caption{Comparison of secure robustness for different methods under uncertainty.}
\label{fig:robustness_vs_uncertainty}
\vspace{-15pt}
\end{figure}
Figure~\ref{fig:robustness_vs_uncertainty} compares the 10th-percentile secrecy spectral efficiency and outage probability for the proposed DRO-CVaR algorithm against all benchmarks, evaluated over three 1500-slot trials. Under the significant uncertainty condition of $\kappa=5$ and $\rho=0.75$, the proposed algorithm achieves 3.18 bit/s/Hz. This represents an improvement of approximately 68\% over the 1.9 bit/s/Hz achieved by the second-best SCA benchmark under identical conditions. The performance gap widens as uncertainty increases, validating the enhanced resilience provided by our distributionally robust approach.

\section{Conclusion}
This letter proposed a two-stage DRO-CVaR framework for robust deployment and beamforming in a secure A-RIS system under multifaceted uncertainties. Our approach decouples long-term UAV placement from per-slot beamforming, using CVaR as a distribution-free risk measure. A low-complexity algorithm was developed, combining a surrogate model for efficient deployment with an AO scheme for robust real-time beamforming. Simulation results validated that the proposed framework significantly enhances the tail-end SSE and maintains the lowest outage probability against benchmarks, proving its effectiveness for providing reliable security in practical A-RIS systems.

\vfill

\end{document}